\documentclass[10pt]{article}
\usepackage{a4wide}                      
\usepackage{epsf}                        
\usepackage{latexsym}                    
\usepackage{amsfonts}                    
\usepackage{amssymb}                     

\newcommand{\sect}[1]{\setcounter{equation}{0}\section{#1}}

\def\be{\begin{equation}}
\def\ee{\end{equation}}
\def\bea{\begin{eqnarray}}
\def\eea{\end{eqnarray}}

\def\nnw{\nonumber \\ [.2cm]}

\def\hsp#1{\hspace*{#1}}

\def\part{\partial}
\def\tfrac#1#2{{\textstyle{\frac{#1}{#2}}}}
\def\half{\tfrac{1}{2}}

\def\tR{{\tilde R}}

\def\cH{{\cal H}}
\def\cK{{\cal K}}  
\def\cL{{\cal L}}
\def\cO{{\cal O}}

\def\mn{{\mu\nu}}
\def\mnr{{\mu\nu\rho}}
\def\mnrl{{\mu\nu\rho\lambda}}

\def\sqrtg{\sqrt{|g|}}

\def\makeatletter{\catcode`\@=11}
\makeatletter
\def\mathbox#1{\hbox{$\m@th#1$}}%
\def\math@ccstyles#1#2#3#4#5#6#7{{\leavevmode
      \setbox0\mathbox{#6#7}%
      \setbox2\mathbox{#4#5}%
      \dimen@ #3%
      \baselineskip\z@\lineskiplimit#1\lineskip\z@
      \vbox{\ialign{##\crcr
             \hfil \kern #2\box2 \hfil\crcr
             \noalign{\kern\dimen@}%
             \hfil\box0\hfil\crcr}}}}
\def\mathaccstyles{\math@ccstyles\maxdimen}
\def\maththroughstyles{\math@ccstyles{-\maxdimen}}
\def\unity%
 {\maththroughstyles{.45\ht0}\z@\displaystyle {\mathchar"006C}\displaystyle 1}
\begin{document}
\begin{titlepage}
\vspace{5mm}
\title{{\bf Palatini versus metric formulation in higher-curvature gravity }} 
\vskip2in
\author{
{\bf M\'onica Borunda\footnote{\baselineskip=16pt E-mail: {\tt
mborunda@ugr.es}}}, \ 
{\bf Bert Janssen\footnote{\baselineskip=16pt E-mail: {\tt
bjanssen@ugr.es}}} \ 
{\bf and} \ 
{\bf Mar Bastero-Gil\footnote{\baselineskip=16pt E-mail: {\tt
mbg@ugr.es}}}  
\hspace{3cm}\\ ~ \\  
{\small  Departamento de F\'{\i}sica Te\'orica y del Cosmos and} \\  
{\small Centro Andaluz de F\'{\i}sica de Part\'{\i}culas Elementales}
 \\  
{\small Universidad de Granada, 18071 Granada, Spain}
}
\date{} 
\maketitle 
\def\baselinestretch{1.15} 
\begin{abstract}
We compare the metric and the Palatini formalism to obtain the Einstein equations in the presence 
of higher-order curvature corrections that consist of contractions of the Riemann tensor, but not 
of its derivatives. We find that there is a class of theories for which the two formalisms are 
equivalent. This class contains the Palatini version of Lovelock theory, but also more Lagrangians
that are not Lovelock, but respect certain symmetries. For the general case, we find that 
imposing the Levi-Civita connection as 
an Ansatz, the Palatini formalism is contained within the metric formalism, in the sense that any 
solution of the former also appears as a solution of the latter, but not necessarily the other way 
around. Finally we give the conditions the solutions of the metric equations should satisfy in 
order to solve the Palatini equations. 
\end{abstract}

\vskip-12.5cm
\rightline{UG-FT-225/08}
\rightline{CAFPE-95/08}
\rightline{May 2008}

\end{titlepage}

\setcounter{footnote}{0} \setcounter{page}{2}

\sect{Introduction}

One of the main lessons of General Relativity is that spacetime is not just a static playground 
in which physics takes place, but that it acts as a dynamical entity, with physical degrees of 
freedom, just like the matter and field content. In General Relativity the gravitational 
interaction is related to the curvature of spacetime, which in turn is determined by the 
energy-momentum content of the spacetime. Free test particles will follow geodesic curves, which 
in general in a curved manifold will not be straight lines. Whereas in Newtonian Mechanics the 
deviation from a straight line is interpreted as a force acting from a distance, in General 
Relativity it is seen as a purely geometrical property of the spacetime. The mathematical picture 
of spacetime that thus arises from General Relativity is that of a $D$-dimensional manifold, 
equipped with a metric $g_\mn$ and a connection $\Gamma_\mn^\rho$, whose dynamics is described by 
the Principle of Minimal Action.  

In differential geometry, the metric and the connection are two independent quantities: the 
former measures distances between points of the manifold and angles between vectors in the tangent 
space, the latter defines parallel transport of vectors and tensors and hence determines the 
intrinsic curvature of the manifold. However assuming the connection to be symmetric 
($\Gamma_\mn^\rho
= \Gamma_{\nu\mu}^\rho$) and metric compatible ($\nabla_\mu g_{\nu\rho}=0$), one finds that the 
connection (and hence the curvature of the manifold) is uniquely determined by the metric 
components:
\be
\Gamma_\mn^\rho = \half g^{\rho\lambda} \Bigl( \part_\mu g_{\lambda\nu} + \part_\mu g_{\mu\lambda}
                                             - \part_\lambda g_\mn \Bigr).
\label{Levi-Civita}
\ee 
This connection is called the Levi-Civita connection and due to its symmetries, the torsion 
({\it i.e}. the antisymmetric part of the connection) vanishes identically and many of the 
curvature tensor identities simplify considerably.

In General Relativity, one usually (tacitly) assumes that the Levi-Civita connection describes 
correctly the physics in Nature and hence that the metric is the only dynamical variable in the 
theory. Besides the uniqueness and simplicity arguments given above, there are also physical 
reasons to prefer the Levi-Civita connection among more general ones. A first reason is that the 
Equivalence Principle, the corner stone of General Relativity which states that the gravitational 
force can be locally gauged away by a convenient choice of coordinates, translates mathematically 
into the ability to set the connection components locally to zero, $\Gamma_\mn^\rho(p) =0$. However, 
due to the tensorial character of the torsion, this is only possible if the connection is 
completely symmetric. 

A second physical reason to prefer the Levi-Civita connection is that for arbitrary connections, 
affine geodesics and metric geodesics do not coincide. Metric geodesics describe the shortest curve 
between two points, while affine geodesics are curves with covariantly  constant tangent vectors. 
Physically the former would represent curves obtained via an action principle, while the latter 
represent unaccelerated trajectories. If both curves do not coincide, it is not clear which one 
would represent the trajectories of free particles, even possibly giving rise to problems with 
causality \cite{ESJ}.

Despite the many arguments in favour of the Levi-Civita connection, it would be nice to have a 
rigorous, mathematical argument to select this one among more general ones. In fact, such an 
argument exists, at least in standard Einstein gravity. In the so-called Palatini formalism 
one assumes that the Ricci tensor in the Einstein-Hilbert action is independent of the 
metric and depends only on the (yet undetermined) connection. The metric appears in fact only in 
the volume element and the contractions of the Ricci tensor, such that the Einstein equation, 
the variation of the action with respect to the metric, is straightforward, albeit involving 
curvature tensors in terms of an arbitrary connection. On the other hand, the variation of the 
action with respect to the connection then yields that the connection should be symmetric and 
metric compatible and hence identifies it uniquely as Levi-Civita. Hence the combination
of the equations of motion of the metric and the connection is physically equivalent to the 
equation of motion coming from the variation of the metric in the action assuming directly the 
Levi-Civita connection \cite{Palatini}. In this sense Levi-Civita arises as a natural solution 
to the Principle of Minimal Action.

Though the Einstein-Hilbert action is the natural choice for an action for gravity in four 
dimensions, there is no reason to exclude higher-order curvature corrections in higher dimensions. 
Already in the 1930's Lanczos \cite{Lanczos} added a specific combination of 
curvature-squared terms, giving rise to a second order, divergence-free, modified Einstein equation 
and in the 1970's Lovelock \cite{Lovelock} generalised this result to arbitrary 
dimensions and arbitrary higher-order curvature terms. In string theory higher-order corrections 
appear naturally as stringy corrections to the supergravity action \cite{CHSW}, and in the 1980's
it was pointed out by Zwiebach \cite{Zwiebach} and Zumino \cite{Zumino} that these stringy 
corrections would bring in ghosts and spoil the consistency of the theory unless the corrections 
appear precisely in the combinations found by Lanczos and Lovelock. Furthermore it was noted in
\cite{Zumino} that the $n$-th order term of this series is precisely the 
$2n$-dimensional Euler character, but elevated to arbitrary dimensions $D>2n$. The relative 
coefficients between the different orders are not determined by first principles, but can be fixed
by the requirement that the theory should have the maximum number of degrees of freedom \cite{TZ}.  

The absence of ghosts make Lovelock gravities the natural extension of Einstein gravity to 
dimensions higher than four. Not all the higher-order corrections that appear in string theory 
are of the Lovelock type \cite{stringy}, but possibly the presence of other fields, in particular 
the dilaton, might cure the problem. Recently higher-order curvature corrections have attracted a 
lot of attention in cosmology, where it is investigated whether these terms could lead to 
corrections to the FRW dynamics that mimic dark matter or cause late-time acceleration 
\cite{cosmologies1}. Likewise their effects have been explored in the context of string and brane 
world cosmology \cite{cosmologies2}. The derivation of the equations of motion of these 
higher-order curvature terms through the metric formalism is an increasingly difficult task, so 
it is a natural question to ask whether the Palatini formalism remains valid in the presence of 
these higher-curvature corrections, especially given its simplicity for the Einstein-Hilbert case.

The Palatini formalism has recently been studied in detail in different contexts, such as $f(R)$ 
gravity, Ricci-squared gravities and comparing the diffeomorphism symmetries of both formalisms 
\cite{ CMQ, ESJ, LBM, Palpapers}. In general the Palatini formalism can lead to more general 
connections than Levi-Civita. In order to compare both formalisms, it is natural therefore to 
impose the Levi-Civita connection as an Ansatz in the Palatini formalism. It was recently pointed 
out \cite{ESJ} that, for a large class of theories in which the Lagrangian 
$\cL (g_\mn, R_\mnr{}^\lambda)$ is a functional of metric and the curvature tensors, but not of its 
derivatives, the Palatini formalism is only equivalent to the metric formalism for those actions 
that describe Lovelock gravities. We will confirm the results of \cite{ESJ} and show that there 
are more Lagrangians for which the two formalisms are  equivalent provided that the Lagrangians 
have sufficient symmetries. In general however, the two formalisms are not equivalent as they lead 
to different equations of motion and we find that in this case the Palatini formalism implies the 
metric formalism for all actions $\cL (g_\mn, R_\mnr{}^\lambda)$, while the opposite is not true 
in general. Solutions of the Palatini equations will always appear as solutions of the metric
equations, but the metric solutions only solve the Palatini equations if they have enough 
symmetries.

The organisation of this letter is as follows: in section \ref{EH} we will briefly discuss 
the Palatini formalism for the Einstein-Hilbert action, basically to review the formalism and 
to set our notation. In section \ref{GB} we will compare the metric formalism and the Palatini 
formalism for the explicit example of quadratic curvature corrections ({\it i.e.} Gauss-Bonnet 
gravity with arbitrary coefficients) and in section \ref{general} we will give a general proof 
for all Lagrangians of the type $\cL (g_\mn, R_\mnr{}^\lambda)$ that the Palatini formalism 
implies the metric one, but not necessarily the other way around. We also study under which 
conditions solutions of one of the formalism also solves the other one. Finally, in section 
\ref{example} we will give a concrete counterexample of a solution of the metric equations that 
does not satisfy the equations of the Palatini formalism and in section \ref{conclusions} we 
summarise our results. 

\sect{The Palatini formalism for Einstein-Hilbert}
\label{EH}

The dynamics of spacetime is traditionally described by the Einstein-Hilbert action, 
\be
S (g)= \int d^Dx \ \sqrtg \ g^\mn R_\mn(g),
\label{Einstein-Hilbert}
\ee 
where $R_\mn(g)$ is the Ricci tensor of the spacetime metric\footnote{Our conventions are as 
   follows: for the metric we use the mostly minus convention, the Riemann tensor is given by 
   $R_\mnr{}^\lambda = 2\part_{[\mu} \Gamma_{\nu]\rho}^\lambda 
          + 2\Gamma_{[\mu|\sigma|}^\lambda \Gamma_{\nu]\rho}^\sigma$, the Ricci tensor by $R_{\mn} = 
   R_{\mu\lambda\nu}{}^\lambda$, the Ricci scalar by $R= g^\mn R_\mn$ and the torsion  by $T_\mn^\rho = 
   2 \Gamma_{[\mn]}^\rho$.} $g_\mn$.
The explicit derivation of the Einstein equations from the Einstein-Hilbert action is rather 
involved, as the action is non-linear and second-order in the metric, such that one would at first 
sight expect
the fields equations to be of fourth order. However, an explicit calculation shows that the 
contribution from the variation of the Ricci tensor vanishes\footnote{The fact that the variation 
of 
     the Ricci tensor does not yield higher-order differential equations is due to the fact 
     that the Einstein-Hilbert action is in fact the first-order term of Lovelock gravity.} 
and only the variation of the volume element and the inverse metric of the Ricci scalar contribute 
to the field equations, yielding the well-known vacuum Einstein equation
\be
R_\mn (g) - \half g_\mn R(g) = 0.
\label{Einsteineqn}
\ee

There exists also a shorter way to obtain the same result, namely the so-called Palatini 
for\-ma\-lism. The formalism consists in assuming that the spacetime manifold is equipped with 
an arbitrary connection $\Gamma_\mn^\rho$, independent of the metric, such that the 
Einstein-Hilbert action is a function of both the metric and the connection, which should now 
be considered as two independent fields, 
\be
S (g, \Gamma)= \int d^Dx \ \sqrtg \ g^\mn R_\mn(\Gamma).
\label{Einstein-Hilbert2}
\ee 
Since the Ricci tensor does no longer depend on the metric, the variation with respect to $g_\mn$ 
gives rise directly to the Einstein equations, 
\be
R_\mn (\Gamma) - \half g_\mn R(\Gamma) = 0,
\label{Einsteineqn2}
\ee
though now the Ricci tensor and scalar are no longer related to the spacetime metric. Yet we should
also vary the action with respect to the connection, the other independent field in the action.
Using the Palatini identity for arbitrary connections
\be
\delta R_\mnr{}^\lambda = \nabla_\mu(\delta \Gamma_{\nu\rho}^\lambda) 
                            - \nabla_\nu(\delta \Gamma_{\mu\rho}^\lambda)
                            + T_\mn^\sigma (\delta \Gamma_{\sigma\rho}^\lambda), 
\ee 
and integrating by parts, the variation of the action (\ref{Einstein-Hilbert2}) is given by  
\bea
0 \ \equiv \ \delta S(g, \Gamma)&=& \int d^D x \sqrtg (\delta \Gamma_\mn^\lambda)
\Bigl[ \nabla_\lambda g^\mn 
        + \Bigl(\half g^{\sigma\tau} \nabla_\lambda g_{\sigma\tau} + T_{\lambda\sigma}^\sigma\Bigr)g^\mn 
\nnw
&&   \hsp{1.6cm}
        - \nabla_\rho g^{\rho\nu}\delta^\mu_\lambda 
        - \Bigl(\half g^{\sigma\tau} \nabla_\rho g_{\sigma\tau}        
                       + T_{\rho\sigma}^\sigma\Bigr)g^{\rho\nu}\delta^\mu_\lambda 
        + g^{\rho\nu}T^\mu_{\rho\lambda}\Bigr].
\label{varEH}
\eea
Since the variation (\ref{varEH}) is proportional to the covariant derivative of the metric and 
the torsion, it is clear that the integral vanishes if the connection is metric-compatible and 
torsionless, in other words: Levi-Civita. Since the Levi-Civita connection is completely determined 
by the metric, as in (\ref{Levi-Civita}), the Einstein equation (\ref{Einsteineqn2}) reduces 
immediately to the Einstein equation (\ref{Einsteineqn}), obtained through the metric formalism.

For standard Einstein gravity the Palatini formalism has two big advantages: a practical one and a 
philosophical one. The practical 
one is that it is much easier to calculate the Einstein equation than in the metric formalism, and 
the philosophical one is that it tells us that the Levi-Civita connection is not just a convenient 
choice, but a physical requirement, a minimum of the action. Let us briefly comment of both issues.

The easiness of calculating (\ref{Einsteineqn2}) in comparison to (\ref{Einsteineqn}) is obvious, 
the most involved part of the Palatini formalism consists in the calculation of the variation 
(\ref{varEH}). However, the conclusion drawn from (\ref{varEH}) remain valid if we couple the
theory to (bosonic) matter by adding a matter Lagrangian $\cL_\phi$, as long as the matter couples 
minimally to 
gravity. In that case, $\cL_\phi$ will contain only factors of the metric, not of the curvature 
tensors and (\ref{varEH}) will not be affected by the presence of $\cL_\phi$. The only modification 
occurs in the Einstein equations, that will get a contribution from the matter Lagrangian in the 
form of the energy-momentum tensor $T_\mn$:
\be
R_\mn (\Gamma) - \half g_\mn R(\Gamma)  = - \kappa T_\mn.
\label{Einsteineqn3}
\ee 
where $\kappa = 8 \pi G_N$. This scenario, however will no longer hold if matter is coupled 
non-minimally or, the case where we are interested in, in the presence of higher-curvature 
corrections. 

The philosophical advantage is probably at least as important as the practical one. In most works on
General Relativity, the Levi-Civita connection is silently assumed, mostly without even questioning 
possible alternatives. Equation (\ref{varEH}) tells us that the Levi-Civita connection is not just a
convenient choice, but appears as a solution to the equations of motion of the connection. Any other
connection would not (necessarily) be a minimum of the action. 

In the next sections we will investigate how far the equivalence between the metric formalism and 
the Palatini formalism remains valid in the presence of higher-order curvature corrections and look 
up to which point the mentioned advantages of the Palatini formalism still hold.

\sect{Comparison for the quadratic curvature corrections}
\label{GB}

As a first test case to compare the metric and the Palatini formalism, we will look at the simplest 
non-trivial example, namely the case of quadratic curvature corrections. 

In the metric formalism, the most general Lagrangian 
that is quadratic in the Riemann tensor and preserves parity is given by 
\be
S = \int d^Dx \sqrtg \ \Bigr[ \alpha R^2 \ + \ 4\beta R_\mn R^\mn \ + \ \gamma R_\mnrl R^\mnrl \Bigr],
\label{Gauss-Bonnet}
\ee 
where $\alpha$, $\beta$ and $\gamma$ are arbitrary constants. For the case where the coefficients
take the values $(\alpha, \beta, \gamma) = (1, -1, 1)$ the action becomes the well-known 
Gauss-Bonnet
term, which is a topological invariant (the Euler character) in four dimensions, but dynamical in 
$D>4$. The easiest way to calculate the Einstein equation is then to write (\ref{Gauss-Bonnet}) 
in terms of the Riemann tensor,
\be
S = \int d^Dx  \sqrtg \ \Bigl[
     \alpha \ g^{\mu\rho}g^{\sigma\alpha}\delta_\lambda^\nu \delta_\beta^\tau
  \ + \ 4\beta \  g^{\mu\sigma}g^{\rho\alpha}\delta_\lambda^\nu \delta_\beta^\tau  
  \ + \gamma \ g^{\mu\sigma} g^{\nu\tau} g^{\rho\alpha} g_{\lambda\beta}
 \Bigl] R_\mnr{}^\lambda R_{\sigma\tau\alpha}{}^\beta,
\label{Gauss-Bonnet2}
\ee
and vary $S$ with respect to the explicit metric and the metrics inside the Riemann tensors. The 
latter is done via the chain rule, using that for the Levi-Civita connection, the variation of the 
Riemann tensor and the Christoffel symbols are given by
\bea
&& \delta R_\mnr{}^\lambda \ = \ \nabla_\mu (\delta \Gamma_{\nu\rho}^\lambda) 
                            - \nabla_\nu (\delta \Gamma_{\mu\rho}^\lambda), \nnw
&& \delta \Gamma_\mn^\rho \ = \ \half g^{\rho\lambda} \Bigl[
            \nabla_\mu (\delta g_{\lambda\nu}) +  \nabla_\nu (\delta g_{\mu\lambda}) 
                          - \nabla_\lambda (\delta g_\mn) \Bigr].
\eea  
The variation of (\ref{Gauss-Bonnet2}) is then given by
\bea
&& \frac{1}{\sqrtg}\frac{\delta S(g)}{\delta g^\mn} \ \equiv \  H_\mn  = \  
   \alpha \Bigl[ 2 \nabla_\mu \part_\nu R - 2 g_\mn \nabla^2 R + 2 R_\mn R - \half g_\mn R^2 \Bigr] 
\label{H=0}                                           \\ [.3cm] 
&& \hsp{1cm}
   + 4\beta \Bigl[ \nabla_\mu \part_\nu R - \nabla^2 R_\mn - \half g_\mn \nabla^2 R  
        - 2 R_{\rho\mu\lambda\nu} R^{\rho\lambda} - \half g_\mn  R_{\rho\lambda}R^{\rho\lambda} \Bigr] \nnw
&& \hsp{.6cm}
   + \gamma  \Bigl[2 \nabla_\mu \part_\nu R - 4 \nabla^2 R_\mn 
                     - 4 R_{\mu\rho}R_\nu{}^\rho
                     - 4  R_{\rho\mn\lambda} R^{\rho\lambda} 
                     + 2 R_{\mu\rho\lambda\sigma} R_\nu{}^{\rho\lambda\sigma}
                     - \half g_\mn R_{\rho\lambda\sigma\tau}R^{\rho\lambda\sigma\tau}  \Bigr]. \nonumber
\eea
Note that for $(\alpha, \beta, \gamma) = (1, -1, 1)$, all the terms that contain derivatives of the 
curvatures cancel out and $H_\mn$ reduces to the so-called Lanczos tensor,
\bea
H_\mn &=& 2RR_\mn  + 4 R_{\rho\mn\lambda} R^{\rho\lambda}  
        + 2 R_{\mu\rho\lambda\sigma} R_\nu{}^{\rho\lambda\sigma}
        - 4 R_{\mu\rho}R_\nu{}^\rho 
\nnw
&& \hsp{1cm}
        - \half g_\mn \Bigr[ R^2 - 4R_{\rho\lambda}R^{\rho\lambda} 
                             + R_{\rho\lambda\sigma\tau}R^{\rho\lambda\sigma\tau}  \Bigr], 
\label{Lanczos}
\eea
which is purely algebraic in the Riemann tensor, such that the Einstein equation 
$H_\mn = - \kappa T_\mn$ contains only second derivatives of the metric. This is due to the fact 
that the Gauss-Bonnet term is in fact the second-order Lovelock gravity Lagrangian.

It is useful to observe that the divergence of $H_\mn$ vanishes, $\nabla_\mu H^\mn = 0$, not only 
for the case of the Lanczos tensor, but for general values of $(\alpha, \beta, \gamma)$. This is 
an attractive property, since it implies that even though in the general case ghosts will appear, 
at least energy conservation is satisfied.

Let us now compare this equation with the corresponding set of equations from the Palatini 
formalism. Here the situation is a bit more complicated: due to the fact that we are dealing
with an arbitrary connection, the curvature tensors do not posses the same symmetry properties
as in the metric formalism and hence there are many possible generalisations of the metric
action (\ref{Gauss-Bonnet}) for arbitrary connections. The most general quadratic curvature 
action that reduces to the action (\ref{Gauss-Bonnet}) when imposing the Levi-Civita connection 
is given by\footnote{We are grateful to Q. Exirifard and M. M. Sheikh-Jabbari for drawing out 
           attention to this point.} 
\bea
S(g, \Gamma) &=& \int d^Dx \sqrtg \ \Bigr[ \alpha R^2 
      \ + \ \beta_1  R_\mn R^{\nu\mu} + 2 \beta_2 R_\mn \tR^{\nu\mu} + \beta_3 \tR_\mn \tR^{\nu\mu} 
 \nnw
  &&  \hsp{3cm}
   \ + \ \beta_4 R_\mn R^\mn + 2\beta_5 R_\mn \tR^\mn +\beta_6 \tR_\mn \tR^\mn \nnw
   &&  \hsp{2.5cm}
        + \ \gamma_1 R_\mnrl R^{\rho\lambda\mu\nu} + \gamma_2 R_\mnrl R^\mnrl 
                       + \gamma_3 R_\mnrl R^{\mu\nu\lambda\rho}
                                   \Bigr].
\label{Gauss-Bonnet3} 
\eea 
The great number of terms is due to the fact that, for arbitrary connections, we can define two 
inequivalent contractions of the Riemann tensor, namely the Ricci tensor $R_\mn$ and the co-Ricci
tensor $\tR_\mu{}^\lambda$
\be
R_{\mu\rho} \equiv 
R_{\mu\lambda\rho}{}^\lambda
\hsp{2cm}
\tR_\mu{}^\lambda \equiv g^{\nu\rho} R_\mnr{}^\lambda.
\ee
Furthermore, in general neither the Ricci tensor nor the co-Ricci tensor  are symmetric, hence the 
six different contractions between both in (\ref{Gauss-Bonnet3}). Note that for the Levi-Civita 
connection, $\tR_\mn$ reduces to the Ricci tensor for Levi-Civita (upto a minus sign). The action 
(\ref{Gauss-Bonnet3}) then reduces to (\ref{Gauss-Bonnet}) upon imposing Levi-Civita and choosing 
$\beta = \sum_{i=1}^6 \beta_i$ and $\gamma = \gamma_1 + \gamma_2 - \gamma_3$.  

For future convenience, it is useful to define the gravitational tensor $\tilde H_\mn$, the 
variation of the action with respect to the metric, and the connection tensor $K_\mn^\lambda$, the 
variation of the action with respect to the connection, as
\bea
\tilde H_\mn &=& \frac{1}{\sqrtg}\frac{\delta S(g, \Gamma)}{\delta g^\mn}, 
\hsp{2cm}
K^\mn_\lambda = \frac{1}{\sqrtg}\frac{\delta S(g, \Gamma)}{\delta \Gamma_\mn^\lambda}.
\eea
The gravitational tensor $\tilde H_\mn$ for the action (\ref{Gauss-Bonnet3}) is then given by
\bea \tilde H_\mn &=& 2 \alpha  R_\mn R 
 \ + \ \beta_1 (R_{\mu\lambda} R^\lambda{}_\nu + R_{\lambda\mu} R_\nu{}^\lambda)
 \ - \ \beta_2 (2 R_{(\nu|\lambda\sigma|\mu)}R^{\lambda\sigma} +\tR^\lambda{}_\mu R_{\nu\lambda}
                           + \tR^\lambda{}_\nu  R_{\mu\lambda}) \nnw
&& + \ 2\beta_3 R_{\lambda(\nu|\rho|\mu)} \tR^{\lambda\rho}
   \ + \  \beta_4 (R_{\mu\lambda} R_\nu{}^\lambda + R_{\lambda\mu} R^\lambda{}_\nu)
   \ - \ \beta_5 (2 R_{(\nu|\lambda\sigma|\mu)} R^{\sigma\lambda} + \tR^\lambda{}_\mu R_{\lambda\nu} 
                               + \tR^\lambda{}_\nu R_{\lambda\mu} ) \nnw
&&   + \ 2\beta_6 R_{\lambda(\nu|\rho|\mu)} \tR^{\rho\lambda} 
    \ + \ \gamma_1 ( R_{\rho\lambda\sigma\mu} R_\nu{}^{\sigma\lambda\rho} 
                          + R_{\rho\mu\sigma\lambda} R^{\sigma\lambda\rho}{}_\nu)
    \ + \ 2 \gamma_3 R_{\mu\rho\sigma\lambda} R_\nu{}^{\rho\lambda\sigma}   \nnw
&& + \ \gamma_2 (R_{\rho\lambda\sigma\mu} R^{\rho\lambda\sigma}{}_\nu
                         \ + 2R_{\mu\lambda\sigma\rho} R_\nu{}^{\lambda\sigma\rho}
                         \ -  R_{\rho\lambda\mu\sigma} R^{\rho\lambda}{}_\nu{}^\sigma)    
  \ - \half g_\mn \cL,
\label{cH}
\eea
where $\cL$ is the Lagrangian that appears in the action (\ref{Gauss-Bonnet3}). On the other hand
the connection tensor $K_\mn^\lambda$ is given by
\bea
K_\mn^\lambda &=& 
2\alpha \Bigr[g_\mn \nabla^\lambda R - \delta^\lambda_\nu \nabla_\mu  R \Bigr]
\ + \ 2\beta_1 \Bigl[\nabla^\lambda R_{\nu\mu} - \delta^\lambda_\nu \nabla^\sigma R_{\sigma\mu} \Bigr]
\nnw
&& - \ 2\beta_2  \Bigl[-\nabla^\lambda \tR_{\nu\mu}
                      +\delta^\lambda_\nu \nabla^\sigma \tR_{\sigma\mu} 
                      + g_\mn \nabla_\sigma R^{\sigma\lambda}
                      - \nabla_\mu R_\nu{}^\lambda \Bigr] \nnw
&& + \ 2\beta_3 \Bigl[ \nabla_\mu \tR_\nu{}^\lambda  - g_\mn \nabla_\sigma \tR^{\sigma\lambda}  \Bigr]
   \ + 2\beta_4 \Bigl[ \nabla^\lambda R_\mn - \delta^\lambda_\nu \nabla^\sigma R_{\mu\sigma}\Bigr] \nnw
&& - \ 2\beta_5 \Bigl[ -\nabla^\lambda \tR_\mn 
                         + \delta_\nu^\lambda \nabla^\sigma \tR_{\mu\sigma}
                         + g_\mn \nabla_\sigma R^{\lambda\sigma}
                         - \nabla_\mu R^\lambda{}_\nu\Bigr] \nnw
&& + 2 \beta_6 \Bigl[-g_\mn \nabla_\sigma \tR^{\lambda\sigma} + \nabla_\mu \tR^\lambda{}_\nu \Bigr]
   \ + \ 2\gamma_1 \nabla^\sigma \Bigl[R^\lambda{}_{\mn\sigma} - R^\lambda{}_{\mu\sigma\nu}\Big] \nnw
&& + 4\gamma_2 \nabla^\sigma R_{\nu\sigma}{}^\lambda{}_\mu
   \ + \ 4 \gamma_3 \nabla^\sigma R_{\nu\sigma\mu}{}^\lambda  
         \ + \ \cO(\nabla^\lambda g_{\mn}) \ + \ \cO(T_\mn^\lambda),
\label{cK}
\eea
where $\cO(\nabla^\lambda g_{\mn})$ and $\cO(T_\mn^\lambda)$ denote terms proportional to the 
covariant derivative of the metric and the torsion respectively, which we will not write down 
explicitly, for reasons that will become clear later on.

At first sight, the equation (\ref{H=0}) is very different from (\ref{cH}) and (\ref{cK}). Note 
that although (\ref{cH}) contains a number of terms of (\ref{H=0}), it lacks, among others, 
the derivatives of the curvature tensor. A direct consequence of this is that the divergence
$\nabla_\mu \tilde H^\mn$ does not vanish, which seems to suggest that the Einstein equation 
in the Palatini formalism, $\tilde H_\mn = -\kappa T_\mn$ is not consistent.

In Ref. \cite{ESJ} the equivalence between the metric and the Palatini formalism is defined 
by demanding $K_\mn^\lambda \equiv 0$ for Levi-Civita ({\it i.e.} demanding that the Levi-Civita 
connection solves the equation of motion of the connection). Here we will use a different 
definition: instead we take the Levi-Civita connection as an Ansatz, substitute it in the 
equations of motion and see whether the remaining equations are equivalent to the Einstein 
equation of the metric formalism.

Imposing the Levi-Civita connection simplifies the expressions for 
$\cH_\mn \equiv \tilde H_\mn|_{\rm Levi-Civita}$ and 
$\cK_\mn^\lambda \equiv K_\mn^\lambda|_{\rm Levi-Civita}$ considerably, 
\bea
\cH_\mn  &=& 2 \alpha  R_\mn R \ 
       + \ 4\beta \Bigl( R_{\mu\rho} R_\nu{}^\rho + R_{\mu\rho\lambda\nu} R^{\rho\lambda} \Bigr)
      \ + \ 2\gamma R_{\mu\rho\lambda\sigma} R_\nu{}^{\rho\lambda\sigma} \nnw
 &&   \hsp{2cm} 
      - \half g_\mn  \Bigr[ \alpha R^2 - \beta R_{\rho\lambda} R^{\rho\lambda} 
                             + \gamma R_{\rho\lambda\sigma\tau} R^{\rho\lambda\sigma\tau}  \Bigr],
\nnw
\cK_{\mn}^\lambda
&=&    2 (\alpha + \beta) g_\mn \nabla^\lambda R   
     \ + \ 4 (\beta + \gamma) \nabla^\lambda R_\mn \nnw
&& \hsp{2cm}
  \ - \ 2 (\alpha + \beta) \delta^\lambda_\nu \nabla_\mu R 
  \ - \ 4 (\beta+ \gamma) \nabla_\mu R_\nu{}^\lambda.
\label{cK2} 
\eea

The point now is to compare the equations obtained in the different formalisms.
As mentioned above, the main difference between $H_\mn$ and $\cH_\mn$ is the presence of second 
derivative terms in the former. It is then natural to try to write the difference between these
two tensors in terms of derivatives of $\cK_{\mn}^\lambda$. Indeed, since
\bea
\nabla_\lambda \cK_{\mn}^\lambda &=& 
-2 (\alpha + \beta) \Bigl( \nabla_\mu\nabla_\nu R - g_\mn \nabla^2 R\Bigr)
           - 4 (\beta + \gamma) \Bigl( \nabla^\lambda \nabla_\mu R_{\nu\lambda} - \nabla^2 R_\mn\Bigr),
\nnw
\nabla^\mu \cK_{\mn}^\lambda &=&  
-2 (\alpha + \beta) \Bigl( \delta_\nu^\lambda \nabla^2 R - \nabla_\nu \nabla^\lambda R\Bigr)
\ -\ 4(\beta + \gamma) \Bigl( \nabla^2 R^\lambda{}_\nu - \nabla^\rho \nabla^\lambda R_{\rho\nu}\Bigr),
\label{dK}
\eea
it is not difficult to see that in this case one can write $H_\mn$ in terms of $\cH_\mn$ and 
$(\nabla \cK)_\mn$ as
\be
H_\mn \ = \ \cH_\mn  \ - \ \half \nabla_\lambda \cK^\lambda_{(\mn)} 
              \  +\ \tfrac{1}{4}g_{\lambda\mu} \nabla^\rho \cK_{\rho\nu}^\lambda
              \  + \ \tfrac{1}{4}g_{\lambda\nu} \nabla^\rho \cK_{\rho\nu}^\lambda.
\label{H=H+K}
\ee
In other words, the Einstein equation in the metric formalism, $H_\mn = -\kappa T_\mn$, can be 
obtained 
through the Palatini formalism via the above combination of the equations of motion of the metric
and the connection. In the same way, energy conservation in the Palatini formalism is guaranteed
on-shell, since $\cH_\mn$ will be divergence-free when the connection tensor vanishes, 
$\cK_{\mn}^\rho = 0$.

But there is more. From (\ref{cK2}) it is clear that for $(\alpha, \beta, \gamma) = (1, -1, 1)$,
the connection tensor $\cK_\mn^\lambda$ is identically zero and hence $H_\mn = \cH_\mn$. Note that 
this is precisely the Gauss-Bonnet case in the metric formalism. In the Palatini formalism 
however, one has to be careful, since it turns out that
not all combinations of $\beta_i$ with $\beta = \sum_{i=1}^6 \beta_i= -1$ in (\ref{Gauss-Bonnet3})
have vanishing connection tensor.  This is a result of the fact that different choices of the 
parameters $\beta_i$ in (\ref{Gauss-Bonnet3}) lead to physically different theories, even though
they all reduce to (\ref{Gauss-Bonnet}) when we impose the Levi-Civita connection.

Explicitly, in terms of the $\beta_i$, formula (\ref{cK2}) takes the form
\bea
\cK_{\mn}^\lambda
&=&    2 (\alpha - \beta_2 + \beta_3 - \beta_5 + \beta_6)  g_\mn \nabla^\lambda R   
     \ + \ 4 (\beta_1 - \beta_2 + \beta_4 - \beta_5 + \gamma) \nabla^\lambda R_\mn \nnw
&& 
  \ - \ 2 (\alpha + \beta_1 - \beta_2 + \beta_4 - \beta_5) \delta^\lambda_\nu \nabla_\mu R 
  \ - \ 4 (-\beta_2 + \beta_3 - \beta_5 + \beta_6 + \gamma) \nabla_\mu R_\nu{}^\lambda.
\eea 
It is then straightforward to see that $\cK_\mn^\lambda$ vanishes for any Lagrangian 
(\ref{Gauss-Bonnet3}) with
\bea
&&\alpha = \gamma \neq 0, 
\hsp{1.5cm}
\beta_1 + \beta_4 = \beta_3 + \beta_6, \hsp{1.5cm}
\alpha + \beta_1 + \beta_4 - \beta_2 - \beta_5 = 0. 
\label{conditions}
\eea
Note that the quadratic Lovelock Lagrangian in the Palatini formalism,
\bea
S(g, \Gamma) 
&=& \int d^Dx \sqrtg \ 
  \left| \begin{array}{ccc}
       g^{\mu_1\nu_1} & ... & g^{\mu_1\nu_4} \\
         \vdots  &         &  \vdots   \\
       g^{\mu_4\nu_1} & ... & g^{\mu_4\nu_4} 
        \end{array} \right| R_{\mu_1\mu_2\nu_1\nu_2}R_{\mu_3\mu_4\nu_3\nu_4} 
\nnw
&=& \int d^Dx \sqrtg \ \Bigr[ R^2 
      \  -  R_\mn R^{\nu\mu} - 2  R_\mn \tR^{\nu\mu} - \tR_\mn \tR^{\nu\mu} 
          + R_{\mn\rho\lambda}R^{\rho\lambda\mn} \Bigr], 
\label{Lovelock2} 
\eea  
indeed satisfies the above requirements, such that we reproduce the results of \cite{ESJ}, that 
for Lovelock gravities the metric and the Palatini formalism are equivalent. However it is also 
clear that there are other combinations of the parameters that satisfy the conditions  
(\ref{conditions}), but are not Lovelock in the sense of (\ref{Lovelock2}). For example the 
combinations 
\bea
\cL_\beta = R_\mn R^\mn + \tR_\mn\tR^\mn,  \hsp{2cm}
\cL_\beta = R_\mn \tR^\mn,
\eea
lead also to vanishing connection tensor and hence equivalence between both formalisms. The reason
why these Lagrangians lead to the same result is because they mimic the symmetries of the Riemann
tensor in the metric formalism
\cite{ESJ2}. In certain sense what we are finding is that the expression (\ref{Lovelock2})
for the definition of Lovelock gravity is too restrictive and that there are more inequivalent 
Lagrangians in the Palatini formalism that lead to divergence-free, second order Einstein 
equations, as originally demanded by Lovelock \cite{Lovelock}.

\sect{More general Lagrangians}
\label{general}

The relation (\ref{H=H+K}) between the gravitational tensors $H_\mn$ and $\cH_\mn$ and the 
connection tensor $\cK_\mn^\rho$ is remarkable. We will show that this is in fact a general result 
for any Lagrangian that contains contractions of the Riemann tensor, but not of its derivatives,
and discuss what this means for the equivalence of the two formalisms.

Let us now derive the above results for the general case where the action is a functional of the 
metric and (contractions of) the Riemann tensor, but not of its derivatives,
\be
S= \int d^D x \sqrtg \ \cL(g_\mn, R_\mnr{}^\lambda).
\ee
The derivation of the equations of motion of the metric and the connection in both the metric 
and the Palatini formalism is completely analogous to the derivation illustrated in section 
\ref{GB}. In the metric formalism the gravitational tensor $H_\mn$ is then given by
\bea
H_\mn &=& \frac{\delta \cL}{\delta g^\mn} \ - \ \half g_\mn \cL  
  \ + \ \half [\nabla_\alpha, \nabla_\beta] 
      \Bigl(\frac{\delta \cL}{\delta R_{\alpha\beta\rho}{}^\lambda} \Bigr) 
                       g_{\rho(\mu} \delta_{\nu)}^\lambda 
\nnw
&-& \half \nabla_\rho \nabla_\alpha \Bigl(\frac{\delta \cL}{\delta R_{\alpha\beta\rho}{}^\lambda} \Bigr)
                 g_{\beta(\mu}\delta_{\nu)}^\lambda 
\ + \ \half \nabla_\rho \nabla_\beta \Bigl(\frac{\delta \cL}{\delta R_{\alpha\beta\rho}{}^\lambda} 
            \Bigr)
                 g_{\alpha(\mu}\delta_{\nu)}^\lambda 
\nnw
&+& \half \nabla^\lambda \nabla_\alpha \Bigl(\frac{\delta \cL}{\delta R_{\alpha\beta\rho}{}^\lambda} 
              \Bigr)
                 g_{\beta(\mu} g_{\nu)\rho}  
\ - \ \half \nabla^\lambda \nabla_\beta \Bigl(\frac{\delta \cL}{\delta R_{\alpha\beta\rho}{}^\lambda} 
             \Bigr)
                 g_{\alpha(\mu} g_{\nu)\rho},  
\label{H}
\eea
while the gravitational tensor $\cH_\mn$ and the connection tensor $\cK^\mn_\lambda$  are of the form
\bea
&& \cH_\mn \ = \ \frac{\delta \cL}{\delta g^\mn} \ - \ \half g_\mn \cL \nnw
&& \cK^{\mu\rho}_\lambda \ = \  \nabla_\nu \Bigl[ 
           \Bigl(\frac{\delta \cL}{\delta R_{\rho\nu\mu}{}^\lambda}\Bigr)
         - \Bigl(\frac{\delta \cL}{\delta R_{\nu\rho\mu}{}^\lambda}\Bigr)  \Bigr],
\label{cH,K}
\eea
where in $\cK^{\mn}_\lambda$ we have already substituted the Levi-Civita Ansatz. Again it is not 
difficult to see that $H_\mn$ and $\cH_\mn$ are related via the general expression
\be
H_\mn = \cH_\mn - \half \nabla_\rho \cK^\rho_{(\mn)} 
               + \half g_{\lambda\mu} \nabla^\rho \cK_{(\nu\rho)}^\lambda
               + \half g_{\lambda\nu} \nabla^\rho \cK_{(\mu\rho)}^\lambda.
\label{H=H+K2}
\ee 
In Ref. \cite{CMQ} the above result has been derived, but through a completely different approach: 
there the Levi-Civita connection was imposed via a Lagrange multiplier, such that the connection is 
not really an independent field. We, in contrast, following the original Palatini philosophy, 
substitute the Levi-Civita connection only as an Ansatz in the connection equation, which really 
appears as a completely independent equation. Yet it is remarkable that both methods yield the 
same results. 

The question now arises whether the two sets of equations are really equivalent, {\it i.e.}, 
whether any 
solution of one set also solves equations of the other set\footnote{One has to be careful when
       comparing explicit solutions of both sets of equations, as the expressions for the metric 
       are only determined up to coordinate transformations. However, as both sets of equations 
       are fully covariant, this should not lead to confusion.}. 
Apart from the cases where $\cK_\mn^\lambda$ is identically zero, as in Lovelock gravity \cite{ESJ},
clearly, formula (\ref{H=H+K}) states that the Palatini formalism is contained within the metric 
formalism: any solution of the equations of motion in the Palatini formalism 
\bea
\cH_\mn = -\kappa T_\mn, \hsp{2cm}
\cK_{\mu\nu}^\lambda = 0,
\label{palatiniEOM}
\eea
is obviously also a solution of the Einstein equation in the metric formalism, which using 
(\ref{H=H+K2}) can be written as
\be
\cH_\mn - \half \nabla_\rho \cK^\rho_{(\mn)} 
               + \half g_{\lambda(\mu} \nabla^\rho \cK_{\nu)\rho}^\lambda
               + \half g_{\lambda(\nu} \nabla^\rho \cK_{\mu)\rho}^\lambda = -\kappa T_\mn.
\label{metricEOM}
\ee
The opposite however is not necessarily true: in a general solution the different terms in the 
left-hand side of (\ref{metricEOM}) will conspire to satisfy the equation, rather than 
spontaneously 
decompose along the lines of (\ref{palatiniEOM}). In next section we will give a explicit example 
of a solution of the metric Einstein equations that does not satisfy the Palatini equations.

A natural question then is to ask under which conditions solutions of the metric formalism also 
solve 
the Palatini equations and what the physical meaning of these conditions is. From 
(\ref{palatiniEOM}) 
we see that a necessary and sufficient condition is that the connection tensor vanishes, while from 
(\ref{cH,K}) we see that $\cK^{\mu\rho}_\lambda$ has the structure of a divergence, 
\be
\cK^{\mu\rho}_\lambda = \nabla_\nu B^{\nu\mu\rho}{}_\lambda.
\label{conserv}
\ee 
The vanishing of $\cK^{\mu\rho}_\lambda$ 
therefore implies a conserved current $B^{\nu\mu\rho}{}_\lambda$, which depends on the Lagrangian 
under consideration. Since $B^{\nu\mu\rho}{}_\lambda$ can be written in terms of contractions of the 
Riemann tensor, the connection equation imposes certain extra symmetry requirements on the metric 
and only those solutions of the metric equations that posses this symmetry are also solutions of 
the Palatini formalism. 

Let us look at some specific examples. For the Einstein-Hilbert action, the vanishing of 
$\cK^{\mu\rho}_\lambda$ is automatically satisfied, as $B^{\nu\mu\rho}{}_\lambda$ is proportional to the 
metric,
\be
\cK^{\mu\rho}_\lambda ({\rm EH})= \nabla_\nu\Bigl[\ 
      g^\mn \delta^\rho_\lambda 
     \ - \  g^{\rho\mu} \delta^\nu_\lambda \ \Bigr]\equiv 0.
\ee 
This is the reason why for standard Einstein gravity, the Palatini formalism is equivalent to the 
metric formalism: the connection equation does not impose any extra condition. A similar expression
can be found for other theories with vanishing connection tensors.
 
For $\cL = R^2$  however, the connection equation demands
\be
\cK^{\mu\rho}_\lambda (R^2)= -2 \nabla_\nu\Bigl[\delta^\rho_\lambda g^{\mn}R 
                                              - \delta^\nu_\lambda g^{\mu\rho}R \Bigr] = 0,
\ee 
which is solved by metrics with constant Ricci scalar $R=\Lambda$. In a similar fashion,
for the Lagrangians
\be
\cL_\beta = R_\mn R^\mn + 2 R_\mn \tR^\mn + \tR_\mn \tR^\mn, \hsp{2cm}
\cL_\gamma =  R_{\mnrl} R^{\mnrl},
\ee
the vanishing of the connection tensor  impose the conditions 
\bea
&& \cK^{\mu\rho}_\lambda ( \cL_\beta) \ = \ 4 \nabla_\nu\Bigl[\
         \delta^\nu_\lambda R^{\rho\mu} + g^{\rho\mu} R^\nu{}_\lambda
       - \delta^\rho_\lambda R^{\mn} - g^{\mn} R^\rho{}_\lambda
\Bigr] = 0, \nnw
&& \cK^{\mu\rho}_\lambda ( \cL_\gamma )
    \ = \ 4 \nabla_\nu R^{\mu\nu\rho}{}_\lambda  = 0,
\eea
which are fulfilled by Einstein spaces and constant curvature spaces ({\it i.e.} maximally 
symmetric spaces) respectively\footnote{Obviously these are not the only solutions. For example,
        locally symmetric spaces, {\it i.e.} spaces with covariant constant Riemann tensor also 
        satisfy the above conditions.}, 
since for these 
spaces the Ricci tensor and the Riemann tensor are proportional to (combinations of) the metric, 
$R_\mn = \Lambda g_\mn$ and $R_\mnrl = \Lambda (g_{\mu\rho}g_{\nu\lambda} - g_{\nu\rho}g_{\mu\lambda})$. 

Note that, although not necessarily the most general solution to the connection condition, the 
solutions presented above are in certain sense the generalisation of the metric {\bf g}${}_{ab}$ 
introduced in \cite{LBM} in the context of Ricci-tensor-squared gravity. It is then straightforward
to generalise the result for arbitrary Lagrangians: for a given Lagrangian $\cL = \cL(g_\mn, 
R_\mnr{}^\lambda)$, a solution of the metric equations is also a solution of the Palatini equations 
if 
\be
\frac{\delta \cL}{\delta R_\mnr{}^\lambda} 
           = \Lambda \ ( g^{\mu\rho}\ \delta^\nu_\lambda \ - \ g^{\nu\rho}\ \delta^\mu_\lambda ).
\ee

\sect{An example: FRW in (pseudo) Gauss-Bonnet gravity}
\label{example}

In this section we will illustrate the inequivalence of both formalism, by 
comparing the cosmological solutions of both formalisms for the case of a Lagrangian that
is Gauss-Bonnet in the metric formalism, but not Lovelock Gauss-Bonnet (\ref{Lovelock2}) 
in the Palatini approach, 
\be
S  = \int d^Dx \sqrtg\ \Bigr[R^2 - 4 R_\mn R^\mn + R_\mnrl R^\mnrl \Bigr].
\ee
Note that indeed this action does not satisfy the conditions (\ref{conditions}) and instead 
imposes the constraint 
\be
\cK_{\mu\rho}^\lambda\Bigr|_{\rm pseudo-Gauss-Bonnet}  \ 
     = \ - 4 \nabla^\lambda G_{\mu\rho}  - 4 \nabla_\mu G_\rho{}^\lambda \ = \ 0,
\label{hier}
\ee
which clearly is solved for metrics that are cosmological vacuum solution $G_\mn = \Lambda g_\mn$, 
with $G_\mn = R_\mn -\half g_\mn R$. 

Consider now an Ansatz of the FRW type
\be
ds^2 = dt^2 - e^{2A}\bar g_{mn} dx^m dx^n
\label{FRW}
\ee
where $A=A(t)$ is a function of $t$ and $\bar g_{mn} = \bar g_{mn}(x)$ is the metric on the spatial 
sections, which for simplicity we assume to be flat, {\it i.e.} $\bar R_{mnp}{}^q =0$. Using 
Eq. (\ref{Lanczos}), 
we find that for the metric (\ref{FRW}), the gravitational tensor of 
the metric formalism is given by  
\bea
\begin{array}{l}
H_{tt} \ =\  -\frac{(D-1)!}{2(D-5)!} (A')^4, \\[.3cm]
H_{mn}\ = \ \frac{(D-2)!}{(D-5)!} \ e^{2A} \bar g_{mn}\Bigl[ 2 A'' (A')^2 + \half (D-1) (A')^4\Bigr],
\end{array}
\eea
where the prime denotes a derivative with respect to $t$.

We will parametrise the energy content by a perfect fluid with pressure $P$, energy density $\rho$ 
and equation of state $P = \omega \rho$, such 
that conservation of energy implies that the energy scales with $A$ as 
\be
\rho = \rho_0 \ e^{-(D-1)(1+\omega)A},
\ee
where $\rho_0$ is the energy density at the initial time $t=t_0$.
In order to solve the Einstein equation $H_\mn = - \kappa T_\mn$, we have to distinguish between
$\omega = -1$ ({\it i.e}. a cosmological constant) and $\omega \neq -1$. For the case $\omega = -1$
the solution is given by de Sitter space
\be
A(t) = \beta t, \hsp{2cm} 
\beta = \Bigl[ \frac{2 (D-5)! \kappa \rho_0}{(D-1)!}\Bigr]^\frac{1}{4},
\label{desitter}
\ee
while for $\omega \neq -1$ we find a power law behaviour
\be
e^{A(t)} = \beta \ \Bigl( \frac{t}{t_0} \Bigr)^{\frac{4}{(D-1)(1+\omega)}},  \hsp{1cm} 
\beta^{(D-1)(1+\omega)}= \frac{(D-1)^3 (D-5)!}{2^7 (D-2)!} \ \kappa \rho_0 (1+ \omega)^4 .
\label{powerlaw}
\ee

In the Palatini formalism, by substituting the Ansatz (\ref{FRW}) in the expressions (\ref{cH}) and 
(\ref{cK2}) for $\cH_\mn$ and $\cK_\mn^\lambda$ obtained in section \ref{GB}, we find that 
\bea
\begin{array}{l}
{\cal H}_{tt}\ =\ - (D-1)(D-2) \Bigl[ 4(A'')^2 + 4 A''(A')^2 + \half (D-3)(D-4) (A')^4 \Bigr],\\
   [.3cm]
{\cal H}_{mn }\ = \ e^{2A} \bar g_{mn}\Bigl[-4(D-2) (A'')^2 + 2(D-2)^2 (D-5) A''(A')^2
                       + \frac{(D-1)!}{2(D-5)!} (A')^4 \Bigr], 
\end{array}
\eea
and
\bea
&&{\cal K}_{tt}^t \ = \ -8 (D-1)(D-2) A'' A', \nnw
&&{\cal K}_{mt}^p \ = \ 8 (D-2) \ \delta_m^p  A'' A', \nnw
&&{\cal K}_{tm}^p \ = \ -4 (D-2) \ \delta_m^p \Bigl[A''' + (D-2) A'' A' \Bigr], \nnw
&& {\cal K}_{mn}^t \ = \ 4 (D-2) \ e^{2A} \bar g_{mn} \Bigl[A''' + (D-2) A'' A'\Bigr]. 
\label{connectionFRW}
\eea 
It is straightforward to see that the de Sitter solution (\ref{desitter}) satisfies the connection 
equation $\cK_\mn^\lambda =0$, as hence also the Einstein equation $\cH_\mn = -\kappa T_\mn$, as 
expected from (\ref{H=H+K}). The power law solution (\ref{powerlaw}) however does not satisfy 
the connection equation, nor the Einstein equation. 

This is indeed what one would expect from the analysis of the previous section. In contrast to the 
de Sitter space, the power law solution does not satisfy the vacuum cosmological Einstein equations
$G_\mn = \Lambda g_\mn$. Or, the other way around, pseudo-Gauss-Bonnet gravity supports FRW type 
solutions only when the matter Lagrangian satisfies the symmetry requirement imposed by the 
connection condition (\ref{hier}), {\it i.e.} if we have a perfect fluid with $\omega = -1$.

\sect{Conclusions}
\label{conclusions}

In this paper we have compared the Einstein equations obtained via the metric formalism with the 
equations obtained via the Palatini formalism for Lagrangians that contain arbitrary contractions 
of the Riemann tensor, but not of its derivatives. We find that for a certain class of theories,
including (but not exclusively) Lovelock gravities, the two formalisms lead to physically 
equivalent sets of equations, when imposing the Levi-Civita connection as an Ansatz, as 
discussed in section~3. In general however, this is no longer the case.

In general, using the Levi-Civita connection either automatically in the metric formalism or 
as an Ansatz in the Palatini formalism, the Einstein equation of the metric formalism can be 
written as a combination of the Einstein equation of the Palatini formalism and the divergence 
of the connection tensor, as in Eq.~(\ref{metricEOM}), but the two sets are not equivalent, in 
the sense that they do not have the same solutions. The Palatini formalism is contained within 
the metric formalism, as any solution of the Palatini equations is also a  solution of the metric 
equations. The opposite, however, is in general not true: only those solutions of the 
metric formalism that have a vanishing connection tensor are also solutions of the Palatini 
equations. Clearly, the set of solutions of the Palatini formalism is a non-trivial subset of the 
solutions of the metric formulation.

As the connection condition in the Palatini formalism has the structure of a conservation law, one 
can interpret this condition as a symmetry requirement that needs to be satisfied by both the 
matter and  gravity Lagrangians in order to become a 
solution of the Palatini equations. In this paper we listed the necessary symmetries for some 
specific theories.

In the introduction we stated that the Palatini formalism in the Einstein-Hilbert action has two 
advantages: the practical advantage was that it was easier to compute the Einstein equations 
assuming that the curvature tensors are independent of the metric, while the philosophical 
advantage was that the Levi-Civita connection is not just a convenient choice, but a physical 
solution to the Principle of Minimal Action. We see now that in general only a small part of 
these advantages remains in the presence of higher-order curvature corrections: although it is 
in principle still possible to compute the metric gravitational tensor $H_\mn$ using the Palatini 
formalism via the relation (\ref{H=H+K}), in practice the effort of computing $H_\mn$ directly 
from (\ref{H}) is not much more than the effort of calculating the connection equation 
(\ref{cH,K}) and its divergences. 
Concerning the philosophical advantage, we had hoped to find a mathematical argument that would 
pick out the Levi-Civita connection among all other possible connections, as a minimum of the 
action. However, strictly speaking, we can only affirm this for a subset of all solutions of the 
(metric) Einstein equations, namely those that are also solutions of the Palatini formalism. For 
the other solutions, we have no arguments to assume that the Levi-Civita connection is more than 
just a (physically perfectly reasonable) choice. 

However for the special class of theories we found, the practical and philosophical advantages 
are as obvious as for the Einstein-Hilbert case. The fact that the connection tensor is 
identically zero, simplifies the Einstein equation and selects the Levi-Civita connection as a 
minimum of the action. It is remarkable that this occurs precisely for Lovelock theories, which 
were already considered a natural extension of standard Einstein gravity due to its divergence-free,
second order equations.

\vspace{1cm}
\noindent
{\bf Acknowledgements}\\
We wish to thank Q. Exirifard and M.M. Sheikh-Jabbari for enlightening discussions and
private communications, and also D. Blas, M. Blau, L. Boubekeur, R. Emparan, J. Pons, 
T. Ort\'{\i}n and P. Townsend for their usefull comments.
The work of M.B. is done as part of the program {\sl Juan de la Cierva} and the work of B.J.~and 
M.B.G.~is done as part of the program {\sl Ram\'on y Cajal}, both of the Ministerio de Educaci\'on 
y Ciencias (M.E.C.) of Spain. The authors are also partially supported by the M.E.C. under 
contract FIS 
2007-63364 and by the Junta de Andaluc\'{\i}a group FQM 101.


\end{document}